**Magnetic and chemical properties of Cr-based films grown on GaAs(001)**

D. H. Mosca, P. C. de Camargo, J. L. Guimarães, W. H. Schreiner

*Departamento de Física –Universidade Federal do Paraná , C. P. 19091, 81531-990 Curitiba PR, Brazil*

A. J. A. de Oliveira, P. E. N. Souza

*Departamento de Física – Universidade Federal de São Carlso, C. P. 676, 13565-905 São Carlos SP, Brazil*

M. Eddrief, V. H. Etgens

*INSP – CNRS, Université Paris VI et VII, 4 Place Jussieu, 75252 Paris Cedex, France*

**Abstract**

We have investigated the magnetic and chemical properties of very thin Cr films, CrAs, and "arsenized" Cr grown by molecular beam epitaxy on GaAs (001), using x-ray photoemission spectroscopy and SQUID magnetometry. Distinct preparation procedures have been used with the purpose to understand the origin of the ferromagnetic signal observed for this system. It results that Ga segregation and chemical reactivity between Ga and Cr have a negligible contribution in the formation of the different thin films. A clear ferromagnetic response even at room temperature suggests the existence of a very thin interfacial layer formed that can eventually be buried during the growth process.

**Introduction**

In the last decade the integration of magnetic materials on semiconductors has been intensively investigated due to the potential applications for spin-based electronics [1-2]. Mono-arsenides integrated on GaAs play an important role since they are ferromagnetic with high Curie temperatures. Recently, *ab initio* theoretical calculations [3] have predicted a ferromagnetic state at room temperature with half-metallic electronic band structure for CrAs with the zinc blend structure. These interesting properties motivated us to investigate the CrAs/GaAs system [4]. Previous results have shown that very thin CrAs epilayers on GaAs(001) have an orthorhombic structure with a magnetic response at room temperature which comes from the strain constraints of the epitaxial growth, causing an expansion of the **b** axis. The crystallisation and chemical interaction at the interface between Cr and GaAs has been explored for different substrate temperatures and film thicknesses [5]. Even at room temperature, diffusion and chemical reactions may occur close to the interface for many transition metal/GaAs systems [6]. The interface characteristics and compound formation have recently been investigated [7-12], but available results are very limited when arsenisation of Cr is considered.

We have studied the magnetic and chemical properties of the Cr/GaAs interface in respect to the different growth procedures. The results have been mainly obtained after X-ray photoemission spectroscopy (XPS) and SQUID magnetometry. In a previous work [8-9] the reactivity between Cr and GaAs was minimized using Cr thin films grown at room temperature on GaAs(001) surfaces cleaned by an Ar+ ion beam. Following a different approach now, our study regard three kinds of samples, prepared at substrate temperatures of 200°C, under different growth procedures and keeping an As-rich environment for incoming Cr atoms at the GaAs surface.

**Experimental**

The samples investigated in this work are chromium thin films, arsenized Cr films, which hereafter are denoted by Cr-As, and CrAs films. All samples were grown using a conventional molecular beam epitaxy system equipped with reflection high-energy electron diffraction (RHEED) and an in situ X-ray photoelectron spectroscopy (XPS) facilities. First an undoped GaAs buffer layer of 1000Å-thick was initially grown on epiready GaAs(001) substrates following standard growth procedure. At the end, an As-rich β(2x4) GaAs surface was stabilized. The sample temperature was carefully reduced in order to avoid the evolution of the sample surface to the c(4x4) As rich surface structure. All three kinds of samples described below were grown on such GaAs surfaces at 200 °C. The Cr thin films were deposited at a growth rate of 4 Å/min using a standard Knudsen cell. The RHEED diagrams observed during growth of the Cr thin film (not shown) initially fade away revealing however, as the thickness increases, a body centered cubic structure characteristic of a three-dimensional growth front. Arsenised Cr films were obtained starting with a 80Å-thick Cr thin films which was annealed under $As_4$ flux during 30 min at 320°C. Finally, CrAs thin films were evaporated using separate Cr and As cells with a growth rate of 8 Å/min under As rich conditions.

In-situ photoemission measurements were performed at different growth stages using unmonochromated Mg $K_\alpha$ (1253.6 eV) radiation and an energy analyzer with an overall resolution of about 0.8 eV. For ex-situ photoemission depth-profile and magnetic measurements, some samples were protected with an amorphous 100 Å-thick ZnSe capping layer deposited at room temperature. This procedure was checked to protect the sample against air exposure for many weeks and the negligible capping layer contribution to the magnetic measurements was confirmed.

Ex-situ photoemission experiments were performed to acquire elemental profiles as well as the species distribution in the overlayer. A commercial VG Microtech ESCA3000 system with a base pressure of $3 \times 10^{-10}$ mbar using unmonochromated Mg $K_\alpha$ radiation and a hemispherical energy analyzer with an overall resolution of about 0.8 eV which was positioned at 45° to the sample surface normal was employed. The nominal thickness of the films was set to 80 Å and 240 Å for Cr and CrAs films, respectively. XPS profiles were obtained by sequential sputtering with argon ions (3 kV, 5 µA) which corresponds to sputter rate of about 2 Å/min for the ZnSe capping layer as well as 5 and 3 Å/min for Cr and CrAs thin films, respectively.

DC magnetic moment measurements were performed with a SQUID (Quantum Design MPMS-5S) magnetometer from 5 K to 350 K for applied magnetic fields up to 50 kOe. Low-field magnetic moment measurements as a function of temperature for different values of applied fields were carried out combining a conventional zero-field cooling (ZFC) warming run, followed by a field cooling (FC) experiment.

**Results and discussion**

**A. Photoemission experiments**

In Figure 1 is shown Ga *2p$_{1/2}$* core level spectra obtained with in-situ experiments for CrAs and Cr thin overlayers grown on GaAs(001). Unreacted Ga line-shape could be detected during CrAs/GaAs(001) interface formation. During Cr/GaAs(001) interface formation, only a small secondary component, positioned at a lower binding energy of 0.8 eV, may be identified in spectrum.

The behaviour of the As profile at the evolving Cr/GaAs interface (cf. Fig. 2 and 3) is significantly different from that of Ga, since the As *3d* and As *3p* core levels do no show appreciable chemical changes induced by initial Cr coverage (for less than 12 Å coverage). Indeed, the core line-shape is not shifted with increasing Cr coverage, no distinct features were resolved, but only a broadening of the full width at half maximum from 1.6 to 1.8 eV. The Cr/GaAs interface has been previously analysed in detail by Weaver et al. [7] with higher resolution and greater surface sensitivity thanks to synchrotron-radiation photoemission experiments. Two different As chemical environments have been evidenced, namely As from GaAs substrate and As from a reacted layer. The As-reacted species were associated with As atoms in a Cr-As bond with energy position energy shifted of 0.25 eV toward the lower binding energy with respect to As emission signal from GaAs substrate. [7]. Unfortunately, the present core-level emission studies cannot distinguish the two As chemical environments due to our total energy resolution of 0.8 eV. We were able however to observe a broadening of the As lineshape which we attribute to a Cr-As phase formation. When increasing Cr coverage (more than 12 Å), the As core-level move toward higher binding energy until the contribution from the Cr *3p* core becomes clearly visible, growing in relative intensity until dominates the As *3d* core line shape (Fig. 2). For a 80 Å nominal Cr thickness, the As *3p* core level is still about 10% of its starting value, while the Ga signal has completely disappear. This attenuated behaviour allowed us to conclude that the Cr/GaAs interface has been buried in a depth greater than the photoemission probe depth (of about 60 Å = 3λ, where λ is the mean free path of 3d and 3p photoelectrons). Further, the shift of the As signal toward higher binding energy is likely due to the presence of a small quantity of As at the surface even at high coverage (As surface-segregated). Above 12 Å-thick, the overlapping between As *3d* and Cr *3p* spectra and the attenuation behaviour of the total emission signal from the Ga-substrate core-line is consistent with the formation of a continuous and uniform Cr overlayer, concomitantly with a body-centred cubic-like structure as observed by RHEED after the early stage regime of the Cr growth.

The spectrum in the topmost right-hand panel of Fig. 2 and 3 shows the post-growth arsenised 80Å-thick Cr film. These results reveal that the binding energy position of As core level does not shift, whereas the intensity ratio between As and Cr signal changes, indicating an As-enrichment of the close to the surface region. In the case of CrAs/GaAs (left-hand panel of Fig. 2 and 3), only a gradual transition of the As core-level from GaAs to CrAs is found, i.e., without evidence of As chemistry during interface formation, in accordance with a well-defined interface observed by HRTEM (see Figure 1 in Ref. 4),

The Cr/GaAs samples have interfaces with a distinct chemical nature as compared to those previously reported by Weaver et al. [7-11]. It is worth noting that the Cr thin films studied by Weaver and coworkers were grown at room temperature on Ga-rich GaAs(001)-c(8x2) surfaces cleaned by $Ar^+$ bombardment. The Cr/GaAs interface of these films exhibits a significant modification under subsequent thermal annealing between 200° C and 320 °C [8,9]. In contrast, the present Cr thin films were grown under As-rich (2x4) surface stabilized just after the GaAs(001) [4,13] epilayer growth at moderate substrate temperature of 200°C and no evidence of Cr in-diffusion is observed up to annealing temperatures of about 220°C [9]. We think that the main reason for the different interfacial chemistry observed by us comes from the different chemical environment found by incoming Cr atoms when they reach the surface. When the surface shows a Ga-rich environment, it becomes unavoidable that the arriving Cr atoms interact, resulting in a segregation of Ga atoms towards the Cr thin film and with a diffusion of Cr atoms towards the GaAs substrate. In such a mixed interfacial layer the formation of Cr-Ga, Cr-As or even Cr-Ga-As compounds can easily occur under a subsequent thermal activation. We have in our study favoured an As-rich surface environment. The moderate substrate temperature of 200 °C promotes an initial bonding between As and the Cr atoms that are impinging the surface. It seems that this procedure leads to an interfacial layer with a distinct chemical and structural nature.

Let's consider now the evolution of the core level photoemission profiles for the three different procedures. Figure 4 shows the evolution of the different core levels for the three samples just after ZnSe protective layer removal. The elemental profiles were obtained after normalization of the photoemission intensities using the atomic sensitivity factors and Shirley background subtraction. The XPS profile for Cr and Cr-As thin films is shown in the top and middle panel of Figure 4, respectively. The bottom panel of Figure 4 shows the XPS profile for the CrAs thin film.

Due to the small thickness of the three samples (thicker film is 240 Å-thick CrAs), we are probing only the substrate after sputtering of about seventy minutes. The positions of the buried interface are in agreement with the nominal thicknesses for 80Å-thick Cr film and 240Å-thick CrAs film. As except, the initial buried interface of Cr has been modified during thermally-assisted arsenisation of Cr film to produce the Cr-As film. As judged by the Ga 2p profile, there is an evidence for Ga in-diffusion into the Cr-overlayer. Concerning the composition of sputtered GaAs substrate, we noted that the GaAs stoichiometry in GaAs is strongly dependent of the mean free path of the photoelectrons. We measured an intensity ratio of about 0.7: 0.3 between As 3s and Ga 2p core-level spectra, whereas Ga-enrichment of the GaAs is found when As 3s / Ga 3d core-level yield is taken into account. According to the mean fee paths of the photoelectrons (of about 6 Å and 20 Å for Ga 2p and Ga 3d, respectively), the Ga accumulation in depth probing photoemission region is probably due to the preferential sputtering of As from GaAs, as substantiated by similar studies with compound semiconductors compounds [14].

At the beginning of the CrAs sputter profile (bottom panel in Figure 4), the Cr and As emission signal suggest a stoichiometric CrAs film despite a small As deficient probably caused by preferential sputtering of As. After sputtering time of ten minutes, the Cr emission decreases monotonically until the GaAs substrate is exposed. For 80-thick Cr thin film, the Cr emission line shows a monotonic reduction with increasing sputter time, whence early growth stage indicates the formation of a uniform Cr thin film. We also noted an evidence of Ga segregation towards the film surface in the both Cr and CrAs thin films. Comparison between As 3s profile for Cr and Cr-As films reveal that there is an effective incorporation of As atoms into the outer monolayers of the Cr film submitted to arsenisation procedure.

It seems that the Cr-As phase formation is not attained by arsenisation despite the thermal assisted (at 320°C) procedure. It is indicating that the CrAs compound formation is obtained by direct reaction of Cr atoms impinging on As-terminated GaAs(001) surface and/or by kinetic Cr-trapping of some As atoms released from the GaAs substrate due to a Cr-Ga chemical reaction observed at the early stage of the Cr nucleation. The growth of CrAs compound using co-deposition, where Cr atoms and As molecules ($As_4$, $As_2$) impinging on GaAs substrate, is succeed because Cr-Ga interaction is probably suppressed by As overpressure at surface substrate.

In summary, the spectroscopic results obtained for the three thin films show that the formation of continuous and uniform overalyers can be obtained. Their intrinsic elemental composition is in quite good agreement with the three different kinds of sample growth procedures investigated by us.

### B. Magnetic measurements

The magnetization as a function of the applied magnetic field for the three samples is shown in Figure 5, for temperatures of 10 K and 300 K. All samples present characteristic hysteresis loops, with clear saturation magnetizations even at the room temperature. The results of Figure 5 present evidence of ferromagnetic response for all samples with moderate coercive field and low remanent magnetization (see insets along the panel).

According to our previous work [4], the saturation magnetization obtained for the 30 Å CrAs films is approximately 1000 emu/cm$^3$, which would correspond to ~ 3 $\mu_B$/Cr for an orthorhombic CrAs, assuming that only chromium atoms contribute to the magnetic moment. This number of ~ 3 $\mu_B$/Cr is the same obtained by Akinaga et al [12]. Indeed, our results are also comparable in what concerns the coercive field of around 200 Oe and remanent magnetization of about 20 % of the saturation magnetization for all the three samples.

It is worth to insist that none of the equilibrium phases of CrAs present ferromagnetic order and chromium thin films are well known itinerant antiferromagnets. Only metastable interfacial compounds or phases can be responsible for the observed ferromagnetic response.

In order to better understand the ferromagnetic behavior, all samples were measured under field cooling (FC) and zero-field cooling (ZFC) conditions. As shown in Figure 6, all three samples present a magnetic irreversibility below 280 K, even for cooling fields above 1 kOe applied in the film plane. However, a ferromagnetic state is observed with an appreciable magnetic moment, even above room temperature.

The normalized magnetization M/Ms is quite similar for CrAs and for Cr-As films. . Like due to ferromagnetic contribution. Additionally to the magnetic irreversibility, a clear peak around 40 K is observed in the FC and ZFC magnetization curves. This peak indicates the presence of frustration processes of the magnetic moments at low temperatures. This strongly suggests the presence of interacting antiferromagnetic clusters. Metallic Cr segregation and antiferromagnetic Cr-based compounds formation could certainly be invoked to explain such a magnetic signal with appreciable magnetic irreversibility. The stronger magnetic irreversibility indicated by a more pronounced bifurcation between FC and ZFC curves for Cr-As thin films supports the assumption of compositional fluctuations. By the way, the strongest magnetic irreversibility for Cr-As films corroborates the photoemission findings, indicating that the arsenisation of Cr film leads to an As-enrichment of only a few Cr atomic plans close to the film surface rather than a bulk Cr-As compound. We believe that compositional fluctuations could occur between CrAs precipitation during formation of the Cr/GaAs buried interface and arsenised Cr-topmost atomic planes formed by arsenisation procedure. However, we would like to point out that only an intrinsic and very thin interfacial layer buried at the interface between Cr and GaAs, could explain the similar behavior that has been observed for the three samples having a rather distinct composition. Our results seems to indicate that the predominant contribution for the magnetism of these thin films arises from a very thin interfacial layer which could intrinsically be formed when Cr atoms interact with the As-rich GaAs surface, forming a new CrAs ferromagnetic phase [4]. A conclusive understanding about the crystalline structure and the magnetic properties of the interfacial layer between Cr and the GaAs substrate remain an open subject. Further studies would be necessary using more sensitive techniques for a buried interface, like for example, the XMCD. Such studies can help to clarify the local order and the chemistry associated with the Cr/GaAs interface.

**Conclusions**

Molecular beam epitaxy samples have been prepared with distinct Cr, arsenized Cr and CrAs compositions over GaAs(001), which displayed a very similar magnetic behaviour. Ferromagnetic as well as magnetic irreversibility features are shown by all three samples. The contact layer resulting from the interaction of the Cr atoms with the As-rich environment supplied by the As rich GaAs(001) surface can be at the origin of this magnetic response. However, no evidence for a Cr-Ga reaction could been found.

**Acknowledgements**

Financial support from CNPq, CAPES/Cofecub, French Program Action Concerte, Nanosciences-Nanotechologies.and FAPESP (grant 03/02804-8) are gratefully acknowledged.

**Figure Captions**

Figure 1 – In situ Ga 2p core-level spectra measured during growth of Cr and CrAs thin films on GaAs. Ga 2p core-level spectrum of the clean surface is also shown. Ga 2p reacted (open circles) and unreacted (solid circles) interface components are shown at coverage of 12 Å Cr.

Figure 2 – Photoemission spectra of As *3d* core-level for CrAs overlayers on GaAs(001) in left-hand panel and for Cr overlayers on GaAs(001) in right-hand panel. In the topmost right-hand panel is also show the spectrum of post growth arsenised 80 Å-thick Cr. Cr *3p* core emission becomes clearly seen at about 1.2 nm overlayer and dominates after 80 Å-thick Cr is reached.

Figure 3 – As *3p* core spectra collected by analogous coverage to those shown in Fig. 2.

Figure 4 – Photoemission profiles obtained for Cr, Cr-As and CrAs thin film samples after background subtraction and normalization of the photoemission intensities. The solid lines is only to guide the eyes.

Figure 5 – In plane magnetization normalized to saturation magnetization as a function of the applied field measured for Cr, Cr-As and CrAs thin film samples. Inner part of the hysteresis loops are shown in the insets along the panel.

Figure 6 – Field cooling (FC) and zero-field cooling (ZFC) magnetic moment curves measured for Cr, Cr-As and CrAs thin films.

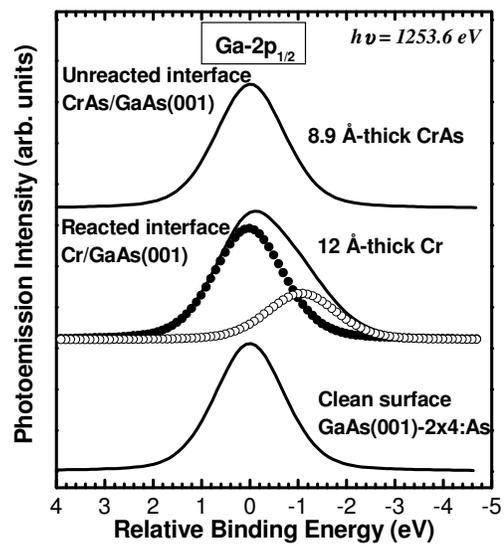

D. H. Mosca et al., Figure 1

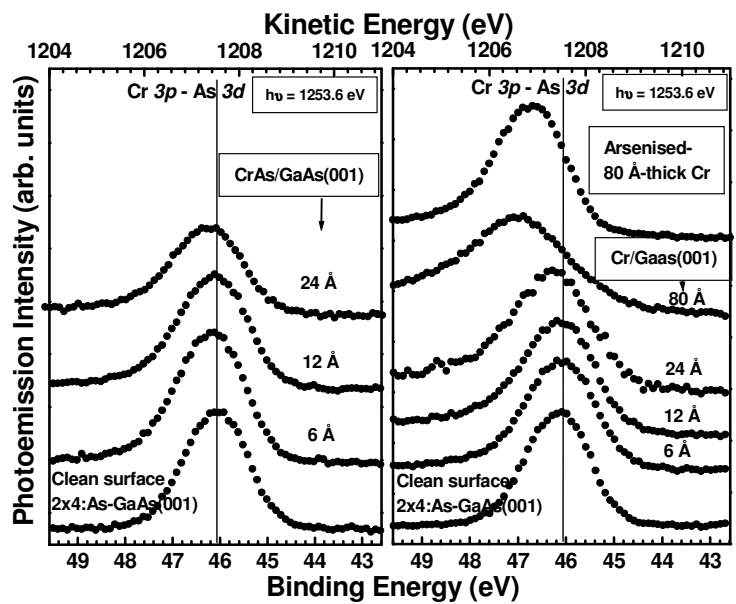

D. H. Mosca et al., Figure 2

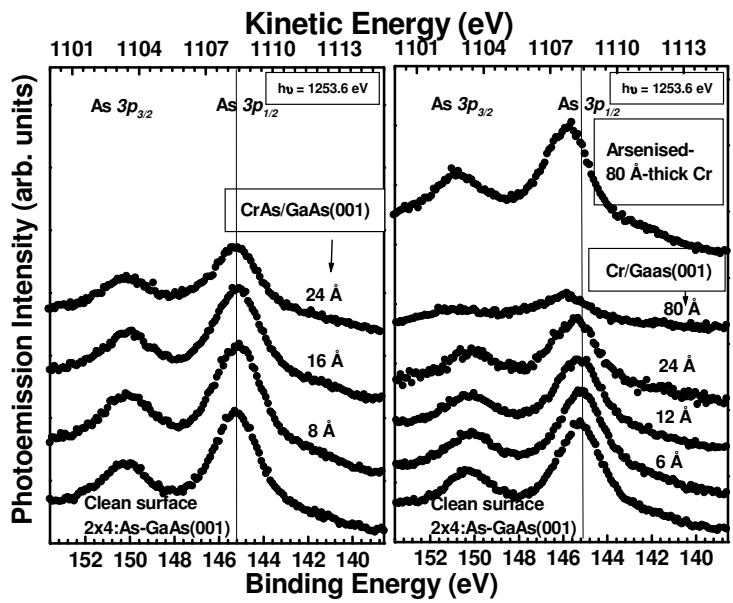

D. H. Mosca et al., Figure 3

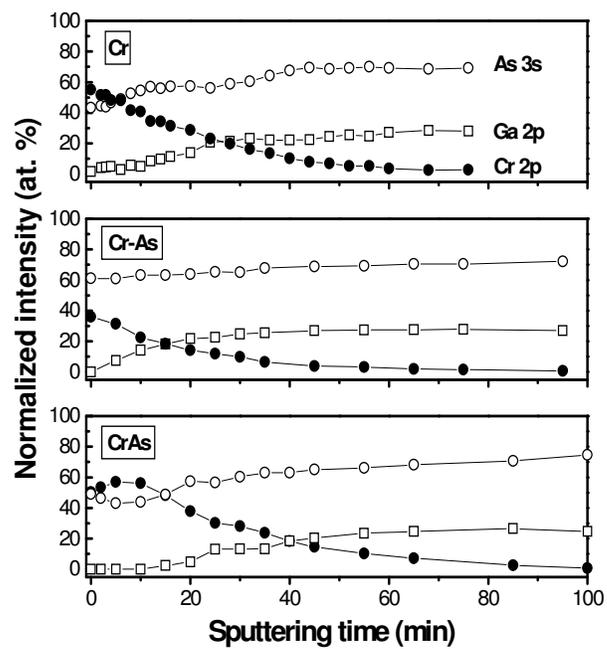

D. H. Mosca et al., Figure 4

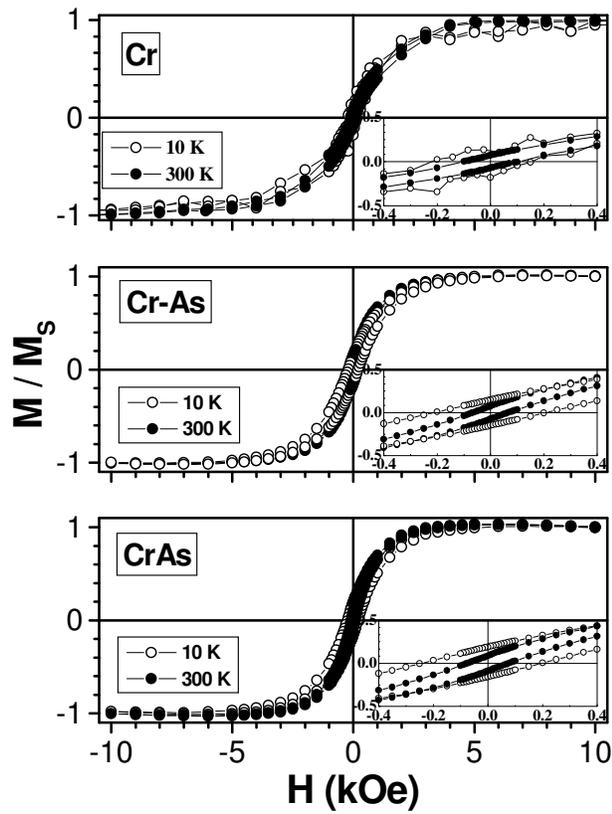

D. H. Mosca et al., Figure 5

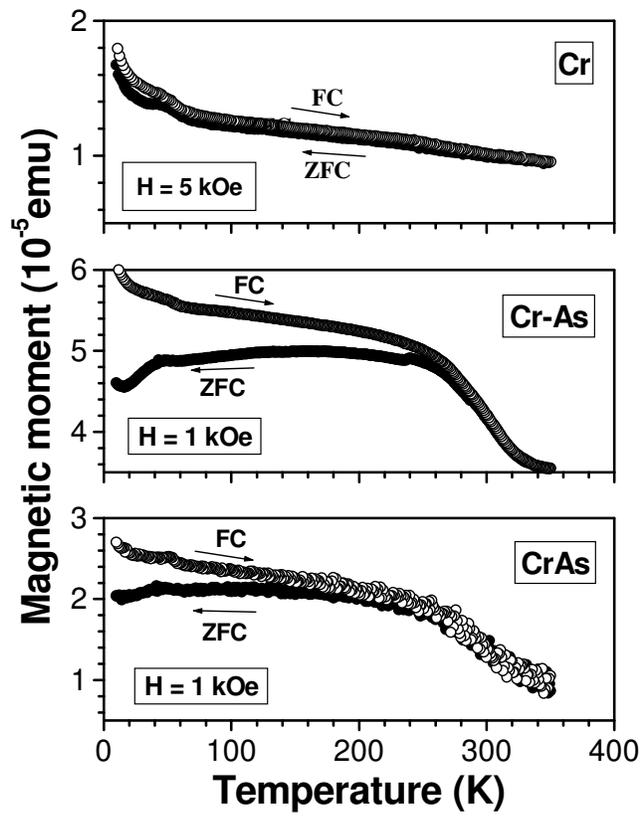

D. H. Mosca et al., Figure 6

16